\documentclass[12pt]{article}
\usepackage{amssymb,amsmath,graphicx,latexsym}
\begin{document}
\title{\textbf{Revisiting $HW\rightarrow\gamma\gamma\ell\nu$}} 
\author{B.~Holdom\thanks{bob.holdom@utoronto.ca}\\
\emph{\small Department of Physics, University of Toronto}\\[-1ex]
\emph{\small Toronto ON Canada M5S1A7}}
\date{}
\maketitle
\begin{abstract}
A new particle of mass around 125 GeV is known to have loop induced couplings to $gg$ and $\gamma\gamma$. More specific to the Higgs interpretation are tree level couplings to $W$ and $Z$. A small but clean signal of the Higgs coupling to $W$ arises from associated production of a Higgs, with $H\rightarrow\gamma\gamma$. We consider this signal in light of a large NLO enhancement of the irreducible background.
\end{abstract}
The cross section for $HW$ production at $\sqrt{s}=8$ TeV is about 700 fb.\footnote{We used the 7 TeV result in \cite{a1} and the ratio of lowest order cross sections at 8 and 7 TeV from MadGraph to scale up the result to 8 TeV.} Multiplying by the standard model branching ratios for $H\rightarrow\gamma\gamma$ and $W\rightarrow \ell\nu$ ($\ell=e\mbox{ or }\mu$) gives about 0.35 fb. Maximum rapidity and isolation cuts on the photons and leptons reduces this to about 0.25 fb. Thus as the integrated luminosity starts to move into the tens of fb$^{-1}$ it becomes feasible to search for these very clean events; namely events with two $\gamma$'s with invariant mass of the Higgs, an isolated lepton and missing energy. This search is sensitive to the same combination of couplings as $W$ boson fusion production of a Higgs decaying to $\gamma\gamma$, but the $WH\rightarrow\ell\nu\gamma\gamma$ cross section is about 10 times smaller \cite{a1}. On the other hand the WBF signal requires stringent cuts to reduce background while $WH\rightarrow\ell\nu\gamma\gamma$ does not; effective cuts can retain most of the signal.

An early study of $WH\rightarrow\ell\nu\gamma\gamma$ can be found in \cite{a8} but no recent theoretical analysis appears to exist. The ATLAS and CMS TDRs \cite{a6,a7} contain studies at $\sqrt{s}=14$ TeV where the irreducible background contributions to $\gamma\gamma\ell\nu$ were considered at lowest order. Recently the NLO QCD corrections to $\gamma\gamma\ell\nu$ production have been obtained including the full leptonic decays of the $W$, that is accounting for the $\gamma$ radiation off the lepton from the decayed $W$ \cite{a3}. (NLO corrections in the stable $W$ approximation were obtained earlier in \cite{a2}).  The $\mathcal{O}(\alpha_s)$ corrections were found to be strikingly large, giving $K\equiv\sigma^{\rm NLO}/\sigma^{\rm LO}$ between 3 and 4 \cite{a3} for cross sections at $\sqrt{s}=14$ TeV, depending on the cuts. This is at least partially due to cancellations at LO as implied by the presence of a zero in the LO amplitude for a particular orientation of the $W$ relative to nearly collinear photons \cite{a4,a10}. We would like to explore the size and effect of the NLO enhancement of the background on the extraction of the $HW$ signal at the LHC.

Most of the NLO enhancement occurs in the tree level amplitudes with an extra final state parton. By inspection of the results in \cite{a3} the loop (virtual) corrections only cause a $\approx25\%$ increase relative to the LO cross section. In addition the full NLO result displays a sizeable sensitivity to the renormalization scale; if the renormalization/factorization scale is one third as large then the full NLO result increases by an amount similar to the virtual contribution. The NLO corrections were also found to modify the shapes of the differential distributions, and we are particularly interested in the phase space region that serves as background to $HW$ production.

The NLO calculation in \cite{a3} is publicly available in VBFNLO \cite{a9} and we shall use this to obtain background predictions. But due to the limited effect of the virtual corrections it also turns out to be useful to study the tree level amplitudes occurring at NLO, as implemented for example in MadGraph (V5) \cite{a5}. A feature of this approach is that the signal and background can be combined at the amplitude level so as to account for interference effects. We also notice that loop corrections enhance the signal cross section by almost 25\% \cite{a1}, and so the missing virtual corrections in both the signal and background in the MadGraph analysis will largely cancel out in $S/B$. Therefore the resulting differential cross section $d\sigma/dm_{\gamma\gamma}$ from this analysis will represent quite well the strength of the Higgs peak above the continuum background.

For the Higgs signal there are, according to MadGraph, 14 independent graphs involving 0 or 1 extra partons contributing to $pp\rightarrow\gamma\gamma e^\pm\nu+X$ tree level amplitudes. (This does not include negligible graphs where the Higgs couples to gluons rather than $W$'s.)  For the irreducible background there are 414 independent graphs (with no Higgs) that contribute to the same final states. The background amplitudes are sensitive to $WW\gamma$ and $WW\gamma\gamma$ vertices and examples of some of the Feynman diagrams can be found in \cite{a10}.

We use the \verb$heft$ model in MadGraph with parameters such that a 125 GeV mass Higgs has a width and branching ratio to $\gamma\gamma$ as given in \cite{a1}. We pass the events through Pythia and PGS. The former is necessary to implement MLM jet matching; we generate 0 and 1 jet samples with the jet $p_{Tmin}$ scale \verb$xqcut=20$ GeV. PGS will illustrate the effect of the electromagnetic calorimeter resolution on the Higgs signal peak. We use
\begin{equation}
\frac{\sigma_E}{E}\approx\frac{10\%}{\sqrt{E}}+0.7\%
\end{equation}
which is representative of ATLAS while CMS has somewhat better resolution. PGS is used with the \verb$LHC$ card and an anti-$k_T$ jet finder with a 0.4 cone size.

We first present results at $\sqrt{s}=8$ TeV. We shall consider two cuts that puts the background into a region of phase space similar to that of the signal. The first is $m_{\gamma\gamma}>100$ GeV and the second is $\hat{H}_T>180$ GeV where the $\hat{H}_T$ is defined as the following scalar sum
\begin{equation}
\hat{H}_T=p_T^{\gamma_1}+p_T^{\gamma_2}+p_T^{\ell}+/\!\!\!p_T
.\end{equation}
The $\hat{H}_T$ distributions and the location of the cut are shown in Fig~1, where the backgrounds have been arbitrarily scaled down in size.
\begin{figure}[t]
\centering\includegraphics[scale=1]{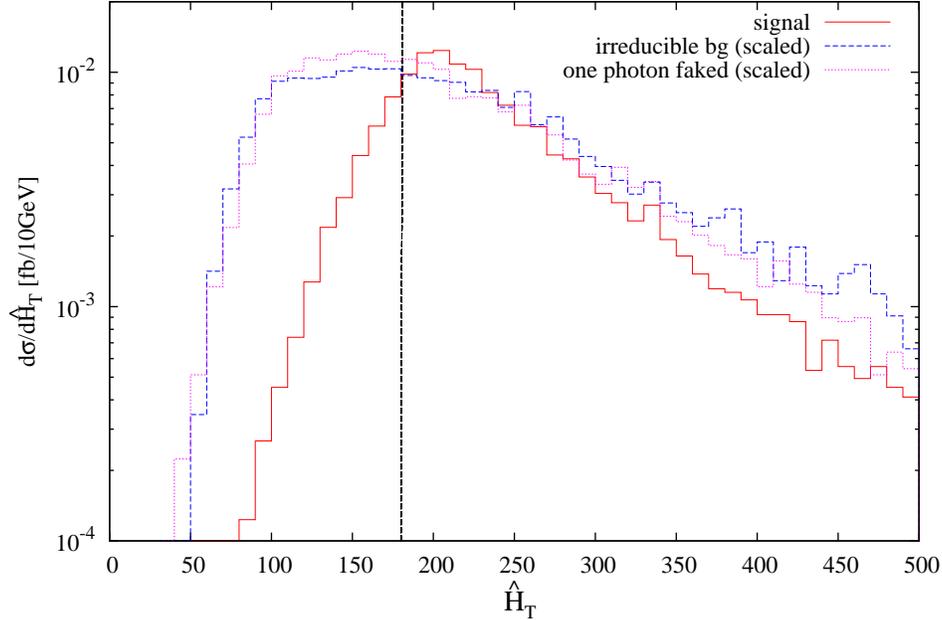}
\caption{Comparing the $\hat{H}_T$ distributions of signal and background}\end{figure}

Our focus is on the irreducible background, but we have also displayed one of the reducible backgrounds, one that arises from $\gamma\ell\nu j$ production where the jet fakes a photon. With the $\hat{H}_T$ cut and with the $m_{\gamma\gamma}>100$ cut or $m_{\gamma j}>100$ GeV cut as appropriate we find that the $\gamma\ell\nu j$ cross section is about 400 times the irreducible background cross section. (The jet in $\gamma\ell\nu j$ is required to have $|y|<2.5$, as for the $\gamma$'s and $\ell$.  We require a $\Delta R>0.4$ separation between any of the reconstructed objects and $p_T>10$ GeV for the $\gamma$'s and $\ell$.) Thus the photon fake rate from jets needs to be less than 1/400 and this is easily achievable.

Other reducible backgrounds without a final state neutrino require the faking of missing energy and are beyond the scope of this work. One example is $e^+e^-\gamma$ production where one of the leptons is reconstructed as a photon \cite{a6}. This can be reduced by a missing energy cut and/or by keeping the invariant mass of any $\gamma e^\pm$ pair away from the $Z$ mass. Such cuts may be imposed without losing much of the signal. The results of  \cite{a6,a7} indicate that this is also true of other reducible backgrounds.

We return to the irreducible $\gamma\gamma\ell\nu+X$ background and obtain the true NLO results from VBFNLO. As a check that the cuts as described have been implemented correctly we compare the LO cross section from VBFNLO to the LO cross section from MadGraph (without an extra parton). We find agreement. For VBFNLO we choose the renormalization/factorization scale $\mu=\sqrt{(p_\ell+p_\nu+p_{\gamma_1}+p_{\gamma_2})^2}$ as in \cite{a3} while for MadGraph we have chosen its default (process dependent) event-by-event  scale choice for $\mu$. We shall also divide these respective scale choices by 3 to probe the scale dependence. The parton distributions functions are \verb$cteq6l1$ for MadGraph and LO VBFNLO and \verb$CT10$ for NLO VBFNLO.

We then use VBFNLO to determine the NLO $K$ factor for the cross section in our region of phase space and find that it ranges from 3.4 to 4 for $\mu$ and $\mu/3$ respectively. The differential $K$ factor $dK/dm_{\gamma\gamma}$ falls not much more than 10\% from 100 to 150 GeV. This large $K$ factor may be surprising when $m_{\gamma\gamma}$ is far from zero, which is where the zero in the LO order amplitude occurs. We find that the $\hat{H}_T$ cut acts to enhance the $K$ factor, and so the NLO enhancement remains large in the phase space region of interest. This suggests that there is more to the large NLO enhancement than the presence of the LO zero.

In Fig.~2 we display our results for the differential cross section $d\sigma/dm_{\gamma\gamma}$ (summing over the $e$ and $\mu$ contributions) in the range between 100 and 150 GeV.\footnote{The programs were run with $95<m_{\gamma\gamma}<160$ GeV.} MadGraph has combined the signal and background at the amplitude level and we see that while lowering the renormalization scale $\mu$ increases the background estimate, the relative signal strength is not diminished. These results also incorporate a PGS acceptance which is about 0.82 for this $m_{\gamma\gamma}$ range, and so we have multiplied the parton level results of VBFNLO by the same factor for comparison purposes. (PGS has also caused a shift in the Higgs peak away from 125 GeV, but we don't expect this to be realistic.)
\begin{figure}[t]
\centering\includegraphics[scale=1]{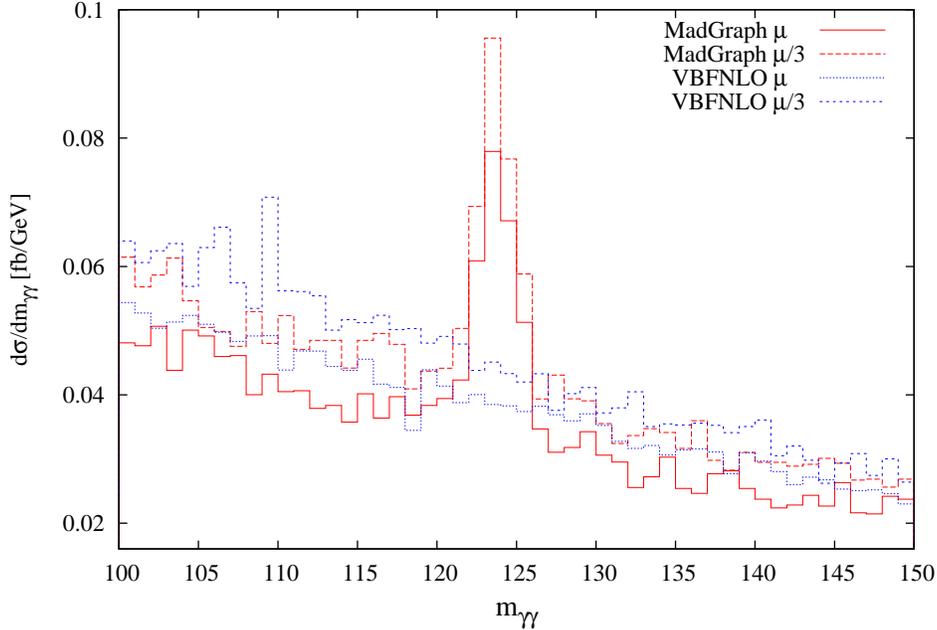}
\caption{$\gamma\gamma$ invariant mass distributions at $\sqrt{s}=8$ TeV. MadGraph includes the Higgs signal while VBFNLO is fully NLO. The renormalization scale $\mu$ is defined differently in the two programs.}\end{figure}
\begin{figure}[t]
\centering\includegraphics[scale=1]{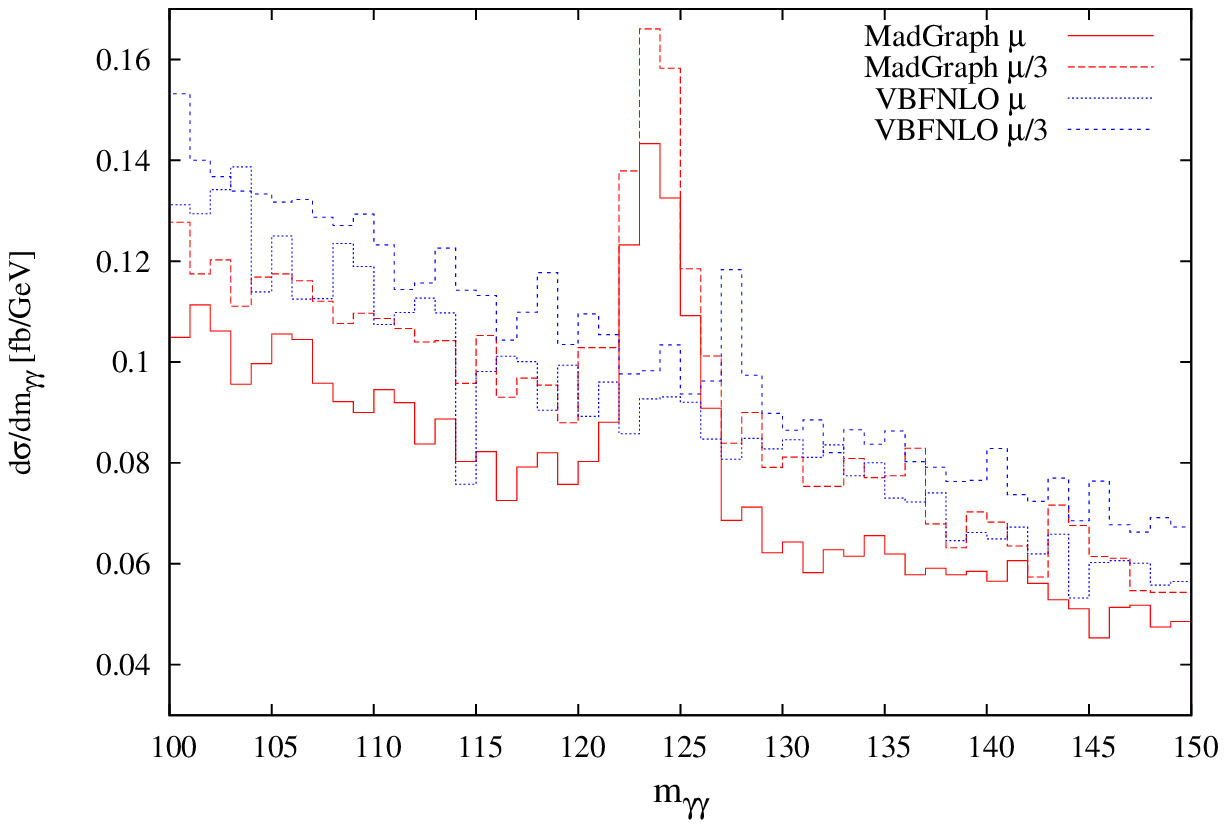}
\caption{Same as previous figure for $\sqrt{s}=14$ TeV.}\end{figure}

The agreement in the background estimates from MadGraph and VBFNLO is good and is consistent with the omission of a  relatively small virtual correction by the MadGraph analysis. This agreement also gives support for the respective choices of renormalization scale in the two programs. The two programs give a similar and quite substantial $\mu$ dependence of the background estimates, but $S/B$ is seen to be quite stable. Finally, we recall that MadGraph has also omitted the virtual correction to the signal, which would act to increase the signal by nearly $25\%$.

 In Fig.~3 we display similar results for $\sqrt{s}=14$ TeV. In this case the LO results from VBFNLO are about 20\% larger than those from MadGraph and we have not corrected for this in the figure. Compared to $\sqrt{s}=8$ TeV the cross sections have about doubled while $S/B$ remains similar. $S/B$ is again stable under change of $\mu$.

We conclude that the process $HW\rightarrow\gamma\gamma\ell\nu$ has an irreducible background that receives very large contributions at NLO, with a $K$ factor as large as 4 according to VBFNLO. Nevertheless we still find a clean signal for this Higgs production and decay mode, showing up as a narrow bump with large $S/B$ at a known location on a smooth background. This is quite unlike other modes that are used to extract the Higgs to $WW$ coupling. The required integrated luminosity is the only drawback; a few events in the signal region should show up for every 10 fb$^{-1}$ collected at $\sqrt{s}=8$ TeV. Not seeing a surplus would provide a useful bound on a product of Higgs couplings. This product is currently allowed to be larger than the standard model value and so an analysis of data already collected could provide a useful constraint.

\section*{Acknowledgments}
I thank Bertrand Brelier for a useful discussion and an anonymous referee for constructive input. This work was supported in part by the Natural Science and Engineering Research Council of Canada.


\begin{thebibliography}{99}
\bibitem{a1} S. Dittmaier et al., LHC Higgs Cross Section Working Group Collaboration, arXiv:1101.0593.
\bibitem{a8} R. Kleiss, Z. Kunszt and W. J. Stirling, Phys. Lett. B253 (1991) 269.
\bibitem{a6} ATLAS Collaboration, G. Aad et al., Expected Performance of the ATLAS Experiment: Detector, Trigger and Physics (2008), CERN-OPEN-2008-020, page 1229, arXiv:0901.0512.
\bibitem{a7} CMS Collaboration, G. l. Bayatian et al., Technical Design Report, Volme II: Physics Performance, CERN LHCC 2006-081, page 318.
\bibitem{a3} G. Bozzi, F. Campanario, M. Rauch, D. Zeppenfeld, Phys.~Rev.~D83 114035 (2011), arXiv:1103.4613.
\bibitem{a2} U. Baur, D. Wackeroth and M. M. Weber, PoS RADCOR2009, 067 (2010), arXiv:1001.2688.
\bibitem{a4} R. W. Brown, K. L. Kowalski, S. J. Brodsky, Phys. Rev. D28 (1983) 624.
\bibitem{a10}  U. Baur, T. Han, N. Kauer, R. Sobey, D. Zeppenfeld, Phys.~Rev.~D56 (1997) 140, hep-ph/9702364.
\bibitem{a9} K. Arnold, J. Bellm, G. Bozzi, M. Brieg, F. Campanario, C. Englert, B. Feigl, J. Frank, T. Figy, F. Geyer, C. Hackstein, V. Hankele, B. Jager, M. Kerner, M. Kubocz, C. Oleari, S. Palmer, S. Platzer, M. Rauch, H. Rzehak, F. Schissler, O. Schlimpert, M. Spannowsky, M. Worek, D. Zeppenfeld, arXiv:1107.4038.
\bibitem{a5} J.~Alwall, M.~Herquet, F.~Maltoni, O.~Mattelaer, T.~Stelzer, JHEP 1106 (2011) 128, arXiv:1106.0522. 

\end{thebibliography}
\end{document}